# Towards a TDD maturity model through an anti-patterns framework

Matheus Marabesi * , Francisco José García-Peñalvo† and Alicia García-Holgado‡
GRIAL Research Group, Universidad de Salamanca (https://ror.org/02f40zc51)
Salamanca, Spain
* https://orcid.org/0000-0001-7646-554X, † https://orcid.org/0000-0001-9987-5584, ‡ https://orcid.org/0000-0001-9663-1103

*Abstract* — *Agile software development has been adopted in the industry to quickly react to business change. Since its inception both academia and industry debate the different shades that agile processes and technical practices play in the day-to-day of students and professional developers. Efforts have been made to understand the pros and cons of the Test Driven Development (TDD) practice to develop software as part of a professional environment. Despite the effort of practitioners to list the TDD anti-patterns that unveil undesired effects in the code when practicing TDD, work is needed to understand the causes that lead to that. In that sense, this paper proposes a research project that explores the TDD anti-patterns context and what leads practitioners to face them in the software development context. As a result, we expect to offer a TDD maturity framework to help practitioners in the process of writing code guided by tests and prevent the addition of anti-patterns.*

*Keywords - TDD, anti-patterns, agile, practitioners, software development.*

I. INTRODUCTION

Software development has changed the way it is done throughout the years, starting with a procedural process something similar to the automotive industry [33] following a well-defined process with a start and end for each stage, once the previous stage is completed then the process will move forward. Bell and Thayer [7] refer to this process as a waterfall.

In the waterfall style, the software process has well-defined phases, gathering requirements, analysis of those requirements, development, testing, delivery and supporting it live. The well-defined process works based on a given context, but it lacks a flexible way to react to business needs. In response to this, the agile manifesto was created [31].

The Agile manifesto was created by practitioners that got together to discuss better ideas to respond to business needs, in that meeting, some ideas came out and the outcome was the manifesto.

There the customer was at the center of the process and the iterative approach was the main focus. Instead of well-defined steps in the process, four statements were used to define the manifesto:

- Individuals and interactions over processes and tools.
- Working software over comprehensive documentation.
- Customer collaboration over contract negotiation.
- Responding to change over following a plan.

Such a shift made the industry and academia change the way they were thinking about software engineering and its process. New ways of working were discovered such as Lean, Kanban and SCRUM [35] as time progressed collaboration was needed going towards a more social activity combined with technical skills.

The iconic book eXtreme programming diffused ideas such as pair programming, feedback and automated tests that were at the time not usual and even preceded the Agile Manifesto [5]. The agile movement and more specifically eXtreme programming brought challenges such as the practice of TDD that practitioners at that time were not used to it.

The value and the practice of writing the test first was not a common approach for software development, as Kent Beck and Erich Gamma noted: "Every programmer knows they should write tests for their code. Few do." [6].

The practices that eXtreme programming brings such as the need for test automation with TDD and design code of quality are the foundation to support an agile environment, that is what practitioners claim when critiquing the SCRUM that focuses on the management side of the software development process. Martin Fowler called that "FlaccidScrum" [17]. The technical practices and more specifically TDD is one practice that allows practitioners to build, iterate and make changes in a controlled way with fast feedback, which in turn might offer the flexibility to respond to business changes, the lack of it, as Martin Fowler noted makes progress slow because the code is a mess.

Still, today the practice of TDD is diffused in the industry, therefore, the results are mixed as Maurício Aniche described in his book [2]:

- TDD made better use of object-oriented programming and decreased 40% to 90% the defects density in comparison with projects that did not use TDD.
- TDD did not accelerate implementation compared with the traditional approach.
- and 14 papers on TDD concluded that TDD shows no consistent effect on internal code quality.

Gustavo Baculi Benato and Plínio Roberto Souza Vilela in their systematic literature review [8] found inconclusive the relationship between cost and benefits that TDD has. Despite its results, TDD is a subject of research and discussion in academia as well as in professional projects. Due to its popularity, there are different styles and interpretations of TDD being used in professional software projects, the two most diffused are outside-in and outside-out [2].

Nevertheless, being a practice that offers benefits to practitioners and is used by high-performing teams as indirectly described through Continuous testing in Accelerate by Jez Humble and Gene Kim [25], what is found in the industry are the misconceptions about TDD, such as: a) replacing the Quality Assurance role, b) when practicing TDD all tests must be written before the actual production code and c) TDD is difficult to learn [3].

In academia, the challenge faced is two folded, the first one being the focus on graduating computer scientists and not software engineers [22] leading to a gap between the academia and the industry.

The second is the way of teaching students the importance of the practice and its benefits, even more, sharing in which context TDD can be used is by itself challenging as it requires professional experience from professors. Last but not least students also perceive testing as a boring activity [29].

In the industry despite the technical skills required to develop software and the collaborative approach described by the agile manifesto, SCRUM which is the most popular framework used focuses on processes [36].

Such popularity has created a gap between technical skills and processes - it is one or the other. In that sense, when a decision is to be made, the one that focuses on process mistakenly wins as the decision makers are not practitioners that are crafting the software, such responsibilities are to be made by Managers, Chief Technology Officers (CTO) as they are the leadership and have the responsibility for the culture in the workplace often related to a not individual contributor role [15] [16].

When an attempt to include technical practices as Mike Cohn did in "Succeeding with agile" [13], the testing strategy besides the automation with TDD is presented as a recommendation named "test pyramid". The pyramid contains its base of unit tests, the middle is integration tests and the top is user interface testing. The shape of the pyramid represents the proportion that each kind of test should have. Therefore, it lacks the context that practitioners face towards achieving the pyramid shape.

Such focus on processes received a pushed back in the form of the craftsmanship movement that started to open the debate on whether such responsibilities should be delegated to a higher level in the hierarchy of the workplace [27], as such focus leads to software projects failing to deliver the expected business value and keep the pace on the long run due to technical debt [18].

Iteratively, focusing on working software (well-crafted software [27]) and collaboration is the foundation of an agile environment. In the literature, we find books that are specific to that [34].

The craftsmanship and the focus on the technical practices are an attempt to educate practitioners and bring attention to their craft - and there are use cases shared showing improvements of quality in the software development cycle [1].

Despite all the work done in the process of software development, the context in which the technical practices are applied such as TDD lacks further investigation to explore the effects that arise when TDD is practiced daily by practitioners.

In that sense, this research project aims to develop a TDD maturity model framework that covers the interactions that lead to the introduction of TDD anti-patterns in the context of software development in an attempt to prevent them.

The structure of this paper is as follows: section II describes the general objectives and the specifics objectives, section III depicts the related work that has been developed, section IV discusses the problem this research aims at, section V describes the methodology planned, section VI describes the work done so far, section VII enumerates the expected results from this research project, and finally, section VIII draw conclusions from the paper presented.

## II. OBJECTIVES

The TDD anti-patterns is a subject that is faced by practitioners, as such, the lack of a structured and defined framework that leads to its cause making the process of solving this problem hidden from practitioner's sight.

In that sense, our objective focuses on tackling the context and technical practices that lead to that. Given that TDD is a practice used by practitioners, the starting point is from the following question:

- Is it possible to prevent the introduction of TDD anti-patterns in software development teams?

Through this broader question, the premises this work proposes are the following:

- Software development team context influences the creation of TDD anti-patterns - The practices that software development teams use on the daily basis might influence how TDD is adopted thus provoking the TDD anti-patterns to arise.

- Technical expertise adds or prevents the addition of TDD anti-patterns - The technical expertise might influence the addition of TDD anti-patterns. The likelihood of experienced engineers adding anti-patterns is lower in comparison to novice engineers.

- The introduction of TDD anti-patterns increases as the lifetime of the application evolves.

Based on those premises, the framework is two-folded. On one hand, the maturity model helps understand which level the practice of TDD is at a given context and code base. On the other hand, given the maturity level, the technical practices also benefit from it allowing each practice to be evaluated to a degree that it adds or prevents the introduction of anti-patterns.

## III. RELATED WORK

More than 10 years ago James Carr came up with a list of TDD anti-patterns to look at and keep under control the pain that practitioners might feel when practicing TDD [12]. The original list that he elaborated on his blog was referenced on StackOverflow [23] containing 22 TDD anti-patterns that are related to the test code itself.

In this section, his list was broken down into four different levels, each level was designed to depict the progress of a practitioner that is starting to learn TDD. Level I is more likely to present issues faced by those just starting to learn TDD. Whereas, IV covers advanced patterns, as the practice of writing tests evolves.

*A. Level I*

- Depending on dependencies such as the operating system can harm testability - The Operating System Evangelist.
- Creating dependencies in which the test runs beyond the operating system can also harm testability (for example, depending on the file system) - The Local Hero.
- Naming test cases are used as a way of debugging and quickly spotting problems, naming them randomly harms understandability - The Enumerator.
- Favor adding new test cases instead of polluting a single test case with many assertions - The Free Ride.
- Avoid coupling test cases with the order in which they appear in a list - The Sequencer.
- While building assertions focus on the specific properties that the test needs instead of comparing an entire object - The Nitpicker.
- Focus on the desired behavior instead of relatively simple actions such as testing a selection from the database – The Dodger.
- Tests that are async-oriented or time-oriented to prevent false positives - The Liar.
- Poluting the test output leads to questioning if the test passed for the right reason - The Loudmouth.

*B. Level II*

- Writing a test that passes first not following the TDD cycle (test failing first) - Success Against All Odds.
- Digging into other object implementations to set up a test case - The Stranger.
- When a test fails and it is difficult to spot the root cause you might be facing a hidden dependency – Hidden Dependency.
- Catching exceptions just to make a test pass - The Greedy Catcher.
- Sharing state between tests whenever possible – The Peeping Tom.
- Relying on exceptions to make the test pass instead make assertions explicit - The Secret Catcher.

*C. Level III*

- Having a test case that does everything at once leading to many lines in a single test case - The Giant.
- Spending too much time setting up the test case points to a code that is not designed for testability, this relates to The mockery - Excessive Setup.
- Violating encapsulation to achieve 100% of code coverage - The Inspector.

*D. Level IV*

- Testing the test double instead of the production code - The Mockery.
- A single test case can have multiple anti-patterns at once - The One.
- Not cleaning up the created data for a specific test case, it is commended to avoid sharing data across tests - Generous Leftovers. This also relates to The Peeping Tom.
- Having a test suite that takes a long time to run – The Slow Poke.

Despite the list that was picked up among practitioners, in 2007 Gerard Meszaros published the xUnit Test Patterns that has a section dedicated to "Test smells" [30]. What he called smells, later became embedded in the "TDD anti-patterns" list.

Martin Fowler [19] also described the pain that practitioners feel when test suites take longer than expected to run or even when tests without any change fail. This scenario is known by academia as a "flaky" test. Martin Fowler elaborates on his scenario using date and time examples. Therefore the flakiness of a test appears in different situations, for example, it appears in the anti-patterns list and relates to "The Peepin Tom".

In 2020 Vladimir Khorikov, dedicated a section of his book to talk about anti-patterns [26] that also relates to what appears in James Carr catalog.

Dave Farley author of Continuous Delivery with Jez Humble [24], went through a few of them on his Youtube channel with an objective point of view and examples from code bases in the open-source community.

Yegor Bugayenko presented a lecture recorded on Youtube about testing patterns and anti-patterns, in his list he summarizes the anti-patterns, besides that he shared with practitioners what he called a "Unit Testing Anti-Patterns — Full List" [11], that combines different sources that are named anti-patterns expanding the list created by James Carr.

IV. THE PROBLEM

The related work presented in the section III depicts different aspects of the reasons that practitioners found to be the reason for their difficulty to write test-first software, it related to both: the source code and the test code. The focus is on the technical aspect of writing code.

Nevertheless, the attention given to the context and how the anti-patterns were introduced lacks further investigation.

In that sense, the adoption of TDD and introduction of TDD anti-patterns might be influenced not only by practitioners that are crafting the code on the daily basis but rather, there is a combination of factors such as:

- practitioner's context that favors learning.
- the perceived added value from the context that practitioners are in such as the stakeholders and technologists.
- the maturity of the team [37].
- the kind of code base practitioner's work: legacy systems [9] or new systems.

Despite the two sides shared here, there is a gap between the processes and the technical practices that do not receive attention in the literature.

Based on the context, anti-patterns can arise when applying TDD decreasing the feedback loop and impacting negatively the perception of the technical practices in a software development context.

## V. METHODOLOGY

This research project relies on Action Research (Fig: 1) as a foundation framework to conduct the activities that aims to explore the practitioner's environment, the emphasis is on what practitioners do [4].

Among the different options to follow a methodology (quantitative, qualitative or mixed) the proposed study aims to analyze the context of practitioners. Such analysis requires a close inspection of how the software development team operates leading to a use-case [32] approach with different groups.

In that sense, the combination of Action Research and a systematic literature review allow us to unveil what has already been developed around TDD and which context TDD is used. Researching what has been done enriches what the current research project is proposing and prevents this project to do what has already been done as well as depict what is lacking further investigation.

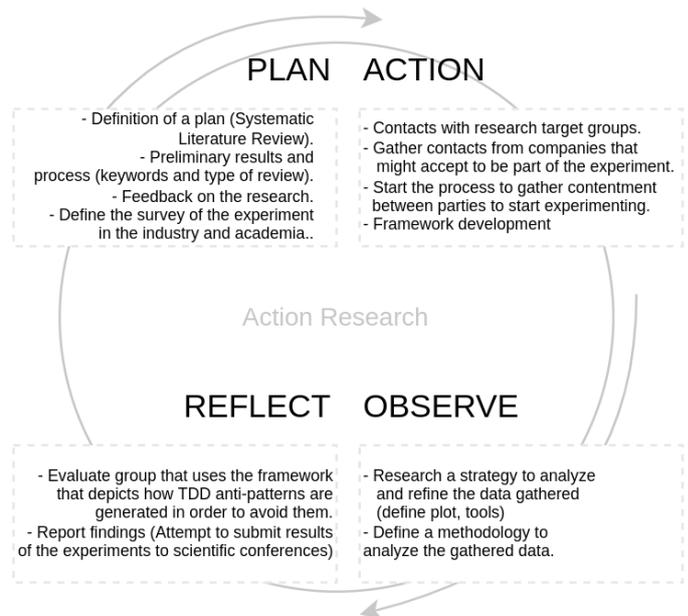

Figure 1. Action Research as a methodological framework. Source: Adapted from [20].

Among the different protocols that can be used to follow a Systematic Literature Review PRISMA (Prefered Reporting Items for Systematic Reviews and Meta-Analysis) will be used, as this is one of the most used for reporting systematic literature review [21].

At first, the qualitative approach will be conducted through an interview as described in the following process:

- Select at least 5 software development groups - The ideal fit for this interview is the team that is working for at least one year together (one year is a guessed number from personal experience, this might be the time to get to the last stage described by [37] to perform in a team) with TDD.
    - from those 5 groups, pick 4 groups and from those 4 select randomly 3 persons.
    - the group remaining will be used to validate the proposed framework.
- For each person that was selected in the group, follow up with an interview to dive into the context in which TDD is practiced. The aim here is to depict not only the technical practices but also the team context in which the TDD is practiced.

Despite the groups being targeted at practitioners in the industry, some places might be worth investigating as a source of data, given that practitioners usually get together to share experiences, launch new products and advance in certain areas that go outside the scientific borders, such as:

- Developers groups in the open source communities that are found in social media such as *Meetup*, *Twitter*, *Facebook* and *Stackoverflow*.
- Practitioners conferences that are not cataloged by journals or academic conferences such as (and not

limited to): *Devoxx, TDD conference, Agile Testing Days, QCon*.

In Action Research, the researcher also wants to evaluate the proposed theory with practitioners and from the feedback improve the theory [4].

In that sense, the framework aims to list comprehensively what are the causes that lead TDD anti-patterns to emerge in code bases, leading to an approach of evaluation in code bases from the mentioned selected groups. The steps are described as follows:

- From the five groups that were used in the previous selection criteria:
    o Pick the remaining group and select randomly 3 persons.
    o Introduction to the framework.
    o Explanation of the steps that the framework aims at preventing TDD anti-patterns.
    o Interview with each participant.
    o Analysis of the data from participants.

Besides that looking at the empirical software engineering arguments that academia started to debate [14], looking at the data in quantitative data would also be beneficial for this research project, for that, different data points can be used to gather insights into other aspects of the effects that the proposed framework might have.

In the context of this project, the data to be collected are mainly from:

- Source code repositories: GitHub (as is one of the most popular platforms for open source projects), GitLab (as it is one of the Github's competitors for open source projects, therefore, it is known for providing private repositories before GitHub.) or any Source Control that uses git.

- Collect the 4 key metrics defined in Accelerate [25] - There are two main possibilities: 1) self-develop a customized tool if needed to fit the research needs, or 2) Collect data through Metrik. Metrik is an open-source tool that automatically collects the 4 key metrics developed by Thought Works.

The collected data from a quantitative fashion allow triangulation of the gathered data leading to a mixed methodology [10].

## VI. THESIS STATUS

The research project proposed is being developed by the Research Group in InterAction and eLearning (GRIAL) at University of Salamanca (Spain). The research group is formed by several researchers from different knowledge fields.

The production that has been done so far includes materials that are focused on the industry as an exploratory approach to get insights from practitioners and responses from software development teams around the subject. The following list depicts the results that came from such exploration:

- A survey in the industry to get insights from what practitioners know about TDD anti-patterns [28].
    o The survey had five sections named: Professional background, TDD practices on the daily basis, TDD practices at companies I worked at, Anti patterns and Finishing up (a section to offer an email to get notified when the data is published).
    o The survey was diffused through Twitter and got 22 answers.
    o The main takeaways from the survey are: a) practitioners learn TDD informally, b) companies from respondents did not require TDD as a skill to join them, c) TDD is not practiced daily and d) The anti-pattern that practitioners recalled the most was The Mockery (further explored in section III).

- A series of talks (six in total) in the software development community - The video series is available on youtube at http://bit.ly/3nJNjhd as a playlist.

The response from practitioners related to the subject revealed a gap that needs development and the thesis is an attempt to formalize such gaps.

The thesis started in 2022 and it is going to be further developed in a timespan of five years and currently, the Systematic Literature Review is under development. The proposed plan in a GANTT fashion to follow is available at http://bit.ly/3YP93Ws for inspection.

## VII. EXPECTED RESULTS

The research aims to study in which context the practice of TDD leads to anti-patterns and starts to become a pain in the daily practice of writing software guided by tests, with that, the following results are expected:

- The first result expected is to bring the discussion about TDD and its anti-patterns when dealing with code bases that have already the practice of writing software with the test-first approach.

- Secondly, as we already presented, in the gray literature practitioners already notice that the practice of TDD does not take into account some aspects of the practice leading to patterns that make testability harder. In that sense, we also would like to contribute with guidance on how to avoid that systematically.

- Last but not least, this project also aims to propose a maturity model to categorize code basis with a maturity model that would help practitioners to improve on the aspects of the TDD anti-patterns.

## VIII. Conclusion

Despite being a popular subject and widely discussed in academia and industry, TDD faces different challenges across its intent to keep as a practice to develop software.

In academia, different approaches were used to evaluate the pros and cons of the practice leading to mixed results. Therefore, in the industry work is still needed to understand what practitioners face when the practice is used but not enough attention is given to the effects that the context might bring.

Furthermore, throughout the design of the methodology, some risks need to be addressed to follow the proposed research project. The following list (that is not exhaustive) depicts such risks:

- The collection of the automated data requires access to source code repositories. It is a common practice to have closed sources for professional groups.
- Adding a constraint in the number of years that the team should be together might lead to a difficult match in the selected groups.
- Due context nature of the project, generalization might not be applied to other groups.

All in all, this research project proposes the development of a framework that will describe the TDD maturity model through anti-patterns in an attempt to prevent the addition of TDD anti-patterns in code bases.